 \documentclass[final,5p,times,twocolumn]{elsarticle}

\usepackage{lineno,hyperref}
\modulolinenumbers[5]

\usepackage{numcompress}

\usepackage{xcolor}

\usepackage{epsfig}

\usepackage{amssymb}

\usepackage{float}

\begin{document}

\begin{frontmatter}



\title{Relativistic approach to a  low perveance high quality matched beam for a high efficiency Ka-Band klystron}

\author{M. Behtouei$^{a}$, B. Spataro$^{a}$, L. Faillace$^{a}$, M. Carillo$^{b}$, A. Leggieri$^{c}$, L. Palumbo$^{b,d}$ and M. Migliorati$^{b,d}$\\\vspace{6pt}}

\address{$^{a}${INFN, Laboratori Nazionali di Frascati, P.O. Box 13, I-00044 Frascati, Italy};\\
 $^{b}${Dipartimento di Scienze di Base e Applicate per l'Ingegneria (SBAI), Sapienza University of Rome, Rome, Italy};\\
 $^{c}${Dipartimento di Ingegneria Elettronica, Universit\`a degli Studi di Roma \textquotedblleft{Tor Vergata}\textquotedblright, Via del Politecnico, 1-00133-Roma, Italia };\\
  $^{d}${INFN/Roma1, Istituto Nazionale di Fisica Nucleare, Piazzale Aldo Moro, 2, 00185, Rome, Italy };\\
  }

\begin{abstract}
Advanced technical solution for the design of a low perveance electron gun with a high quality beam dedicated to high power Ka-band klystrons is presented in this paper. The proposed electron gun can be used to feed linear accelerating structures at 36 GHz with an estimated  input power of 20 MW, thus achieving an effective accelerating electric field in the (100-150) MV/m range. additionally, in the framework of the Compact Light XLS project, a short Ka-band accelerating structure providing an integrated voltage of at least 15 MV, has been proposed for bunch- phase linearization. For the klystron, a very small beam dimension is needed and the presented electron gun responds to this requirement.  An estimate of the rotational velocity at beam edge indicates that the diamagnetic field due to rotational currents  are small compared to the longitudinal volume. A detailed analysis of how this is arrived at, by compression of the beam, rotation in the magnetic field, and analysis of the subsequently generated diamagnetic field has been discussed.

\end{abstract}


\end{frontmatter}



\section{Introduction}

The new generation of high-gradient linear accelerators is extremely demanding in order to produce high-brightness electron beams. These beams are used in a great number of applications, including future linear colliders, X-ray Free-Electron Lasers (FELs) and inverse Compton scattering accelerators for research, compact or portable devices for radiotherapy, mobile cargo inspections and security, biology, energy, and environmental applications.

 In the framework of the \textquotedblleft{Compact Light XLS}\textquotedblright \ project \cite{CL}, a short ultra-high gradient linearizer working on the third harmonic ($\sim$36 GHz) of the bunched electron beam and generated by a high-voltage DC gun (up to 480 kV) operating with an accelerating gradient of $\sim$150 MV/m (i.e., 15-20  MV integrated voltage range) is requested \cite{Behtouei1,Behtouei2,Behtouei3,Behtouei35} . To meet these requirements, a 36 GHz pulsed Ka-band RF power source with a pulse length of 100 ns and a repetition frequency  in the (1-10) Hz range with an output power of (50-60) MW is necessary \cite{Behtouei4,Behtouei5,Behtouei6}.

The klystron amplifier design requires a proper choice of some parameters: perveance, beam and pipe diameters, focusing magnetic field, bunching cavities and output cavity system, ultra-vacuum system, coupling coefficient, plasma frequency reduction factor, and beam collector. Among them, the perveance which is one of the challenging aspects of the high power klystron design has a key role in designing the electron gun. The lower the perveance, the
weaker the space charge, and, consequently, the stronger the bunching. On the other hand, higher perveance causes strong space charge leading to low efficiency because of weak bunching. As a result, finding an optimal perveance to maintain a good efficiency is always a challenging point in electron gun design.
 In this paper, we present an electron gun in the Ka-band with a focusing magnetic device producing a beam radius of 1 mm with the minimum scalloping of 0.98, and confined in a 1.2 mm beam pipe in order to maximize the klystron efficiency. The reason why we kept the scalloping effect within 2$\%$ is that it is an optimized number for the klystron efficiency \cite{Caryotakis}. We show that, with a proper focusing magnetic field, we could manage to minimize the scalloping effect for increasing the coupling parameters.  The electron gun geometry is optimized to adjust the electric field equipotential lines for obtaining an extracted beam current of 100 A. Estimations have been obtained by using  the numerical code  CST Particle Studio \cite{CST} and analytical approach. The analytical results for calculation of the electron gun dimensions have been compared with numerical estimations.

\section{Design procedure of the electron gun and focusing magnetic field}

The main design parameters of an electron gun and focusing magnetic field demand

1) To find an optimal perveance: Perveance, $K = I \ /\  V^{3/2}$ is the parameter by which we control and measure the space charge force. I and V stand for the beam current and voltage, respectively. The higher the perveance, the lower efficiency and vice versa. We have chosen a low micro-perveance of  0.3 $A V^{-3/2}$ for our device so that we have a high efficiency. In the following section it will be demonstrated that the difference between relativistic current density and Child-Langmuir (non-relativistic regime) is small enough that we can consider the non-relativistic approach for calculating the perveance.

 2) Define an optimal electrostatic beam compression ratio and the maximum electric field on the focusing electrode: by solving the Poisson's equation in spherical coordinates and with the help of electrostatic lens effect which is a bridge between light and charged-particle optics, we can find the potential distribution between cathode and anode and consequently it is possible to optimize the geometry of the electron gun in order to have a high electrostatic beam compression ratio and a low electric field strength on the focusing electrode. The electrostatic beam compression ratio has been chosen to be 1500:1 and the maximum electric field on the focusing electrode is about 200 $kV/cm$. The procedure  for estimating the dimensions of the electron gun device is as follows:

 \begin{figure*}[t]

 \begin{center}
\includegraphics[width=0.3 \textwidth ]{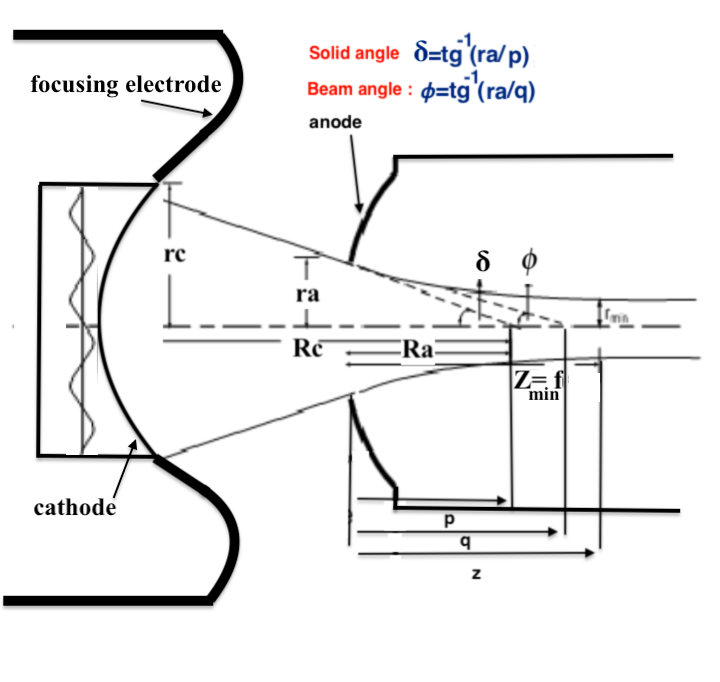}(a)
  \includegraphics[width=0.3 \textwidth ]{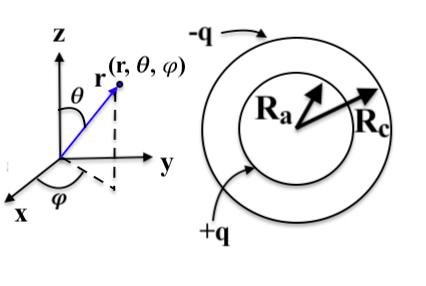}(b)
   \includegraphics[width=0.6 \textwidth ]{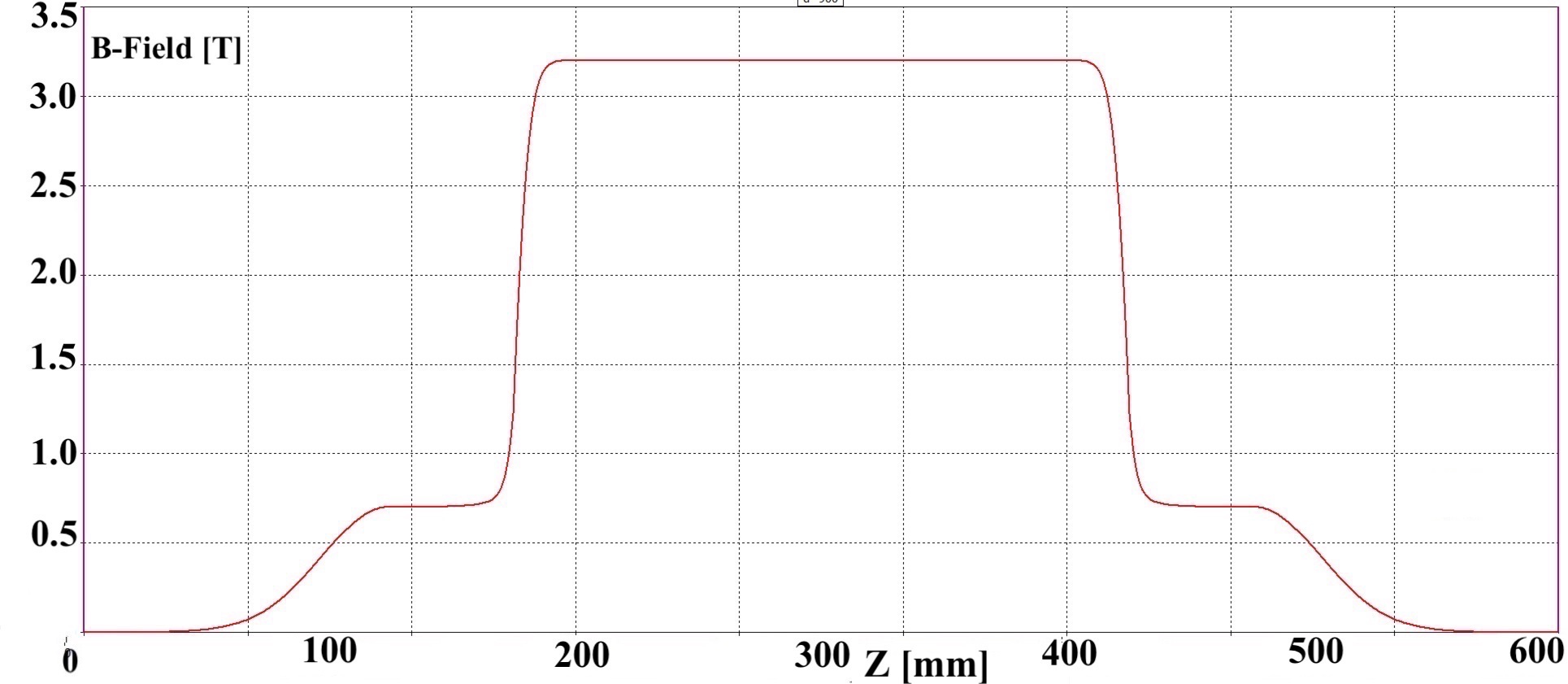}(c)
    \includegraphics[width= 0.6\textwidth ]{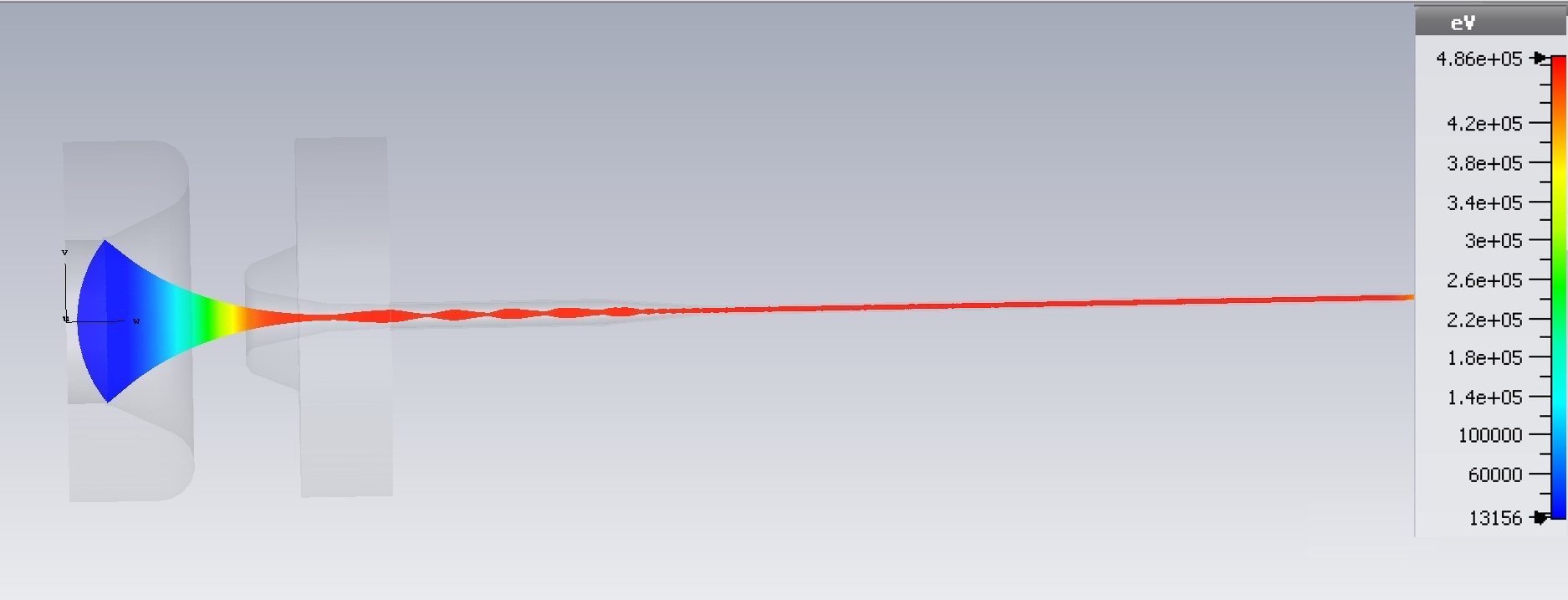}(d)

\caption { a) Schematic view of the Pierce-type gun geometry  b) Two concentric conducting spheres of inner and outer radii $R_a$ and $R_c$, equivalent with the DC gun c) 3D model of the gun and of the beam pipe. The cathode-anode voltage is 480 kV and the beam current 100 A . Beam trajectory along the propagation direction  d)  the axial magnetic field distribution. }

 \end{center}

      \end{figure*}

Poisson's equation in spherical coordinates is given

\begin{equation}
\small{\frac{1}{r^2} \frac{\partial}{\partial r} (r^2 \frac{\partial V}{\partial r})=-\frac{\rho}{\epsilon_0}=\frac{I}{4\pi r^2 \nu \epsilon_0}}
\end{equation}
where $\nu$ is the electron velocity, related to the voltage V by $\frac{1}{2} m_e \nu^2=e V$ with $m_e$ the electron mass and e its charge. The final solution of the above equation is given by \cite{Langmuir,Blodgett},

\begin{equation}\label{4}
\small{I=\frac{4 \epsilon_0}{9} (\frac{-2e}{m_e})^{1/2} \frac{V^{3/2}}{(-\alpha)^2}}
\end{equation}

where,

\begin{equation}\label{5}
\small{\alpha= \xi-0.3\ \xi^2+0.075\ \xi^3-0.0143\ \xi^4+....}
\end{equation}

and  $\xi=log(\frac{R_a}{R_c})$  where $R_a$ and $R_c$ are the radii of the spheres of anode and cathode, respectively  (see Fig. (1b)) \cite{Pierce}.

From Eq. ($\ref{4}$), the beam current is proportional to 3/2 power of the cathode voltage and the constant of proportionality is the perveance. The other parameters, like the beam angle $\phi$ ($\phi=tg^{-1}(r_a/q)$) and anode aperture radius can be obtained from electrostatic lens effect by considering analogy between light and charged-particle optics  \cite{Haimson} (see Fig.(1a) and (1b)). The aperture lens can be obtained \cite{IP}

\begin{equation}\label{6}
\frac{1}{F}=\frac{E_b-E_a}{2 p v}
\end{equation}

where F is the aperture number, $E_b-E_a$ stands for the difference of the energy gradients,  p and v are the momentum and velocity of the charged particles. The higher the aperture number, F, the smaller the aperture hole gets.

3) Define the minimum beam radius in magnetic system: to avoid the increase of the transverse dimensions of the beam inside the beam pipe, after the electron gun exit,  due to the existence of space charge, a transverse focusing magnet is needed. The magnetic field profile and the corresponding beam envelope are shown in Fig.  (1c) and (1d). We reported the design parameters of the gun with the focusing magnetic field along the beam axis in Table 1. It should be noted that the maximum possible beam compression (minimum beam radius) is necessary to avoid the voltage breakdown \cite{Yakovlev}. By minimizing the beam radius, the transverse emittance rises and beam scalloping will start to increase. Through an appropriate focusing magnet system, we obtain an scalloping effect within 2$\%$ which is an optimal number for the klystron efficiency \cite{Caryotakis}. It should be noted that, by having a proper beam dimension, we will get the optimum interaction with the output RF cavities. 
  
 \begin{table}[h]

\caption{Design parameters of the gun with the focusing magnetic field along the beam axis}

 \begin{center}
\begin{tabular}{||c|c|c||} 
\hline
Design parameters&  \\ 
\hline\hline
Beam power [MW]&  48\\ 
\hline
Beam voltage [kV]& 480\\  
\hline
Beam current [A]  & 100\\ 
\hline
 Micro-perveance $[I/V^{3/2}]$&0.3\\  
 \hline
 Cathode diameter [mm]&  76 \\ 
 \hline
  Pulse duration [$\mu$ sec]& 0.1 \\ 
  \hline
Minimum beam radius in magnetic system [mm]& 0.98\\ 
\hline
Nominal radius [mm]& 1.00\\ 
\hline
 Max EF on focusing electrode [kV/cm]&  200 \\ 
 \hline
 Electrostatic compression ratio& 1500:1\\ 
 \hline
Beam compression ratio& 1700:1\\ 
\hline
 Emission cathode current density [$A/cm^2$]&2.02\\ 
 \hline
Transverse Emittance of the beam [mrad-cm]& 1.23 $\pi$\\ 
\hline
\end{tabular}
\end{center}
\end{table}

In Table 2, we compared the analytical and numerical results (Eqs. (\ref{5}) and (\ref{6})) for estimating the dimensions of the electron gun device obtaining a good agreement.

 \begin{table}[h]
\caption{Comparison among analytical and numerical results for estimating the dimensions of the electron gun device}
\begin{center}
\small{\begin{tabular}{||c|c|c||} 
\hline
Parameters& Analytical& Numerical (CST) \\ 
 \hline\hline
$\frac{R_c}{R_a}$&  3.20&3.15\\ 
\hline
$r_a$ [mm]&12&13 \\ 
\hline
Solid angle, $\delta$& 16$^\circ$&18$^\circ$\\ 
\hline
Beam angle, $\phi=tg^{-1} (r_a/q)$&13$^\circ$ &14$^\circ$\\ 
\hline
 Beam current [A]  & 100&100\\ [0.5ex] 
 \hline
\end{tabular}}
\end{center}
\end{table}

4) Define the opportune magnetostatic beam compression ratio: The magnetic field profile, shown in Fig. (1c), has two peaks of 7 kG and 32 kG. The bigger one is located along a distance of about 300 mm to obtain a narrow beam radius to allow cavities to operate on the third harmonic of the fundamental mode of X-band. These cavities work in the Ka-band regime and therefore they require a small beam radius of about 1 mm. The magnetostatic beam compression ratio of 1700:1 is located in this region. It should be noted that the maximum possible compression ratio occurs when the beam radius reaches to the Brillouin limit \cite{Yakovlev}. In the next section we will show that the beam radius of 1 mm is far away from the Brillouin limit and the diamagnetic field effect is negligible.

We will also show why we can use the non-relativistic approach to calculate perveance and how we can neglect the diamagnetic field effect due to the rotation in the magnetic field.

\section{Relativistic approach to the design}

As we mentioned in the previous section, to get a high efficiency klystron, we design a low perveance electron gun. In this section we will show that a non-relativistic approach for calculating the perveance is reliable as the difference between relativistic current density and Child-Langmuir limit current density (non-relativistic regime) is negligible. Then, it will be demonstrated that also the diamagnetic field effect is negligible due to the fact that the rotation frequency is much smaller than Larmor frequency and also the beam radius is far away from the Brillouin limit. Finally we will achieve the minimum scalloping effect for increasing the coupling parameters by combining the beam propagation with a proper focusing magnetic field\cite{Chin}.

It is known that Brillouin flow is produced when the sum of the radial forces on the electrons constituting the stream becomes continuously equal to zero. For this condition the following equation must be satisfied along the stream \cite{Birdsall}

\begin{equation}\label{1}
4 \omega_L \omega_r- 2\omega_r^2=\omega_p^2
\end{equation}

where

\begin{equation}\label{2}
\omega_L =\frac{-B e}{2m_e}\ \ \ \ , \ \ \ \  \omega_r = \omega_L  (1-\frac{\psi_c}{\psi})
\end{equation}

and

\begin{equation}\label{3}
\omega_p^2=-\frac{I e}{\pi m_e \epsilon r^2} \sqrt{\frac{-m_e}{2eV}}
\end{equation}
$\omega_L$ is called the Larmor frequency, $\omega_r$ is called the cyclotron frequency, $\omega_p$ is called the plasma frequency, B is the axial flux density of the magnetic field, e is the electron charge (negative), $m_e$ is the electron mass, I is the current in the electron stream, V is the stream voltage, $\psi_c$ and $\psi$  are the magnetic fluxes linking the helical path of an electron at two points (departure point and Brilliouin radius, respectively) along the stream and r is the equilibrium radius of the beam.

It is desirable to make the cyclotron frequency equal to the Larmor frequency in producing Brilouin flow. Then Eq. ($\ref{1}$) becomes \cite{Rosenzweig1,Rosenzweig2}

\begin{equation}
\omega_p^2=2\omega_L^2
\end{equation}

Substituting Eq. ($\ref{2}$) and Eq. ($\ref{3}$) into the above equation we obtain  equilibrium radius of the beam as

\begin{equation}
 r^2=\frac{-2m_eI}{e\pi  \epsilon B^2} \sqrt{\frac{-m_e}{2eV}}
\end{equation}

The author of \cite{Yakovlev} estimated the Brillouin limit radius with the following equation 

\begin{equation}
\small{r_b=\frac{0.369}{B}\ \sqrt{\frac{I}{\beta\gamma} }\ mm}
\end{equation}

that is derived from envelope equation in space charge domain, with the effect of an external magnetic field.

In our case is I=100A, $\beta=v/c=0.860$, $\gamma$ is the relativistic factor $\gamma=1.939$ and B is magnetic field in kG  B=32 kG. The Brillouin limit radius is about $r_b=$0.1\ mm.

In the region where magnetic field is 32 kG, the beam radius is $\sim$ 1 mm, considerably higher than Brillouin limit which is about 0.1 mm. Likewise for the region where the field is 7 kG, the beam radius is $\sim$ 2.2 mm which again is much bigger than the Brillouin limit which is about 0.4 mm. 

We have also obtained the same limit for the beam radius by considering Caryotakis approach \cite{Caryotakis} and obtained the same result.

Therefore in both cases (B=7kG and 32kG)  the beam radius is far away from the Brillouin limit and the diamagnetic field effect is negligible.

The next step is to calculate relativistic current density and compare it with the non-relativistic approach (Child-Langmuir regime) to get an idea of what is the difference between the relativistic perveance and the non relativistic one. To accomplish this task we define the relativistic perveance as $K = I_{relativistic}/ V^{3/2}$. The relativistic solution of the one-dimensional planar diode has been performed in \cite{Greenwood} 

\begin{equation}
J_{relativistic}=\frac{4}{9} \epsilon_0 \sqrt{\frac{2e}{m_e} \frac{V_0^{3/2}}{d^2}} \ \ \  {\bigg[}\  _2F_1 {\big(}1/4,3/4;7/4;-\frac{V_0 e}{2 m_e c^2}{\big)}{\bigg ]}^2
\end{equation}
where $\ _2F_1 {\big(}1/4,3/4;7/4;-\frac{V_0 e}{2 m_e c^2}{\big)}$ is the ordinary hypergeometric functions having a general form of the kind

\begin{equation}
 \ _pF_q(a_1,..,a_p;b_1,..,b_q;z)=\Sigma_{n=0}^\infty \frac{(a_1)_n ... (b_p)}{(b_1)_n...(b_p)_n}\frac{z^n}{n!}
\end{equation}
where $(a_p)_n, (b_q)_n$ are the rising factorial or Pochhammer symbol with

\begin{equation}
(a)_0=1
\end{equation}

and

\begin{equation}
(a)_n=a(a+1)(a+2)...(a+n-1),\ \ \ \ n=1,2,....
\end{equation}

In the case of non-relativistic regime we have  \cite{Langmuir,Blodgett} 

\begin{equation}
J_{Child-Langmuir}=\frac{4}{9} \epsilon_0 \sqrt{\frac{2e}{m_e} \frac{V_0^{3/2}}{d^2}} .
\end{equation}

In Fig.2 we have shown the ratio between relativistic regime and non-relativistic one as a function of $\frac{V_0 e}{2 m_e c^2}$. 

  \begin{figure}[h]
 \begin{center}
 \fbox{\includegraphics[scale=0.4]{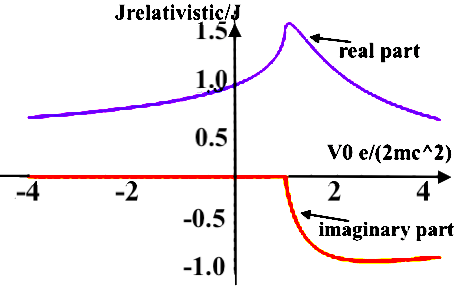}}
\caption{ Real and imaginary part of the ratio between relativistic current density and Child-Langmuir (non-relativistic regime) one. }
\end{center}
 
 \end{figure}

For the cathode voltage of 480kV, the hypergeometric function becomes $_2F_1 (1/4,3/4;7/4;-0.47)=0.96$ and the ratio between the relativistic current density and Child-Langmuir ones becomes about 0.92 ($\frac{J_{relativistic}}{J_{Child-Langmuir}}$=0.92). As Perveance is defined   $I =K V^{-1.5}$,  we have demonstrated that the difference between non-relativistic  and  relativistic regimes is less than 8$\%$ and we can consider the non-relativistic approach for calculating the perveance.

The final investigation is dedicated to see how we can mitigate possible diamagnetic effects. The cyclotron frequency is determined according to the Busch's theorem

\begin{equation}
\omega_r=\omega_L  (1-\frac{\psi_c}{\psi})
\end{equation}

where $\psi_c$ and $\psi$  are the magnetic fluxes linking the helical path of an electron at two points (departure point and Brilliouin radius, respectively) along the stream. This equation, in the case of uniform axial flux density, can be written

 \begin{equation}
\frac{\omega_r}{\omega_L }=1-\frac{r_c^2}{r^2}
\end{equation}

where $r_c$ and r are the radius of the electron path at the departure point and at the second mentioned point, respectively. In our case, with the minimun scalloping, $r_c=0.98$ and r=1, the obtained rotation frequency is $\omega_r=0.04\ \omega_L$ and it means that the rotation frequency is much smaller than Larmor frequency and this mitigate the possible diamagnetic effects.

 \section{Conclusions}
In this paper, we presented a design of an electron gun with a low perveance in order to get a high efficiency klystron. With a proper focusing magnetic field, we managed to minimize the scalloping effect for increasing the coupling parameters. We have shown that the beam radius is far away from the Brillouin limit and the rotation frequency is much smaller than Larmor frequency and this mitigate the possible diamagnetic field effect. Finally we have demonstrated that the difference between relativistic current density and Child-Langmuir (non-relativistic regime) is less than 8$\%$ and we can consider the non-relativistic approach for calculating the perveance. 

The proposed design has allowed the development of a high beam quality class electron gun dedicated  to high efficiency klystrons that are suitable for next generation of harmonic klystron currently under design at the INFN.

\section*{Acknowledgment}
This work was partially supported by the Compact Light XLS Project, funded by the European Union's Horizon 2020 research and innovation program under grant agreement No. 777431 and by INFN National Committee V through the ARYA project.

\end{document}